# QTIA, a 2.5 or 10 Gbps 4-Channel Array Optical Receiver ASIC in a 65 nm CMOS Technology

H. Sun,[a,b,1] X. Huang,[b] C.-P. Chao,[c] S.-W. Chen,[c] B. Deng,[d] D. Gong,[b,2] S. Hou,[e] G. Huang,[a] S. Kulis,[f] C.-Y. Li,[c] C. Liu,[b] T. Liu,[b] P. Moreira,[f] Q. Sun,[g] J. Ye,[b] L. Zhang,[a,b,1] W. Zhang [a,b,1]

[a] *Central China Normal University,*
 *Wuhan, Hubei 430079, PR China*

[b] *Southern Methodist University,*
 *Dallas, TX 75275, USA*

[c] *APAC Opto Electronics Inc.,*
 *Hsinchu, 303, Taiwan*

[d] *Hubei Polytechnic University,*
 *Huangshi, Hubei 435003, China*

[e] *Academia Sinica,*
 *Nangang, Taipei 11529, Taiwan*

[f] *CERN,*
 *1211 Geneva 23, Switzerland*

[g] *Fermi National Accelerator Laboratory,*
 *Batavia, IL 60510, USA*

 *E-mail*: dgong@smu.edu

ABSTRACT: The Quad transimpedance and limiting amplifier (QTIA) is a 4-channel array optical receiver ASIC, developed using a 65 nm CMOS process. It is configurable between the bit rate of 2.56 Gbps and 10 Gbps per channel. QTIA offers careful matching to both GaAs and InGaAs photodiodes. At this R&D stage, each channel has a different biasing scheme to the photodiode for optimal coupling. A charge pump is implemented in one channel to provide a higher reverse bias voltage, which is especially important to mitigate radiation effects on the photodiodes. The circuit functions of QTIA successfully passed the lab tests with GaAs photodiodes.

KEYWORDS: Front-end electronics for detector readout; Optical detector readout concepts; Radiation-hard electronics; VLSI circuits.

---

[1] Visiting scholars at SMU and performed this work at SMU.
[2] Corresponding author.

**Contents**

## 1. Introduction

High-speed optical data communication is widely used for on-detector readout electronics in high-energy physics (HEP) experiments. In the optical receiver end, radiation-induced degradation of photodiode is one of the most challenging issues. Many studies on photodiodes in radiation, mostly by the European Organization for Nuclear Research (CERN), indicate that radiation causes different degradations in GaAs and InGaAs photodiodes [1,2,3]. For InGaAs photodiodes, the dark current increases from nanoampere levels to hundreds of microamperes, and the junction capacitance at low reverse voltages (below 2 V) increases from hundreds of femtofarads to several picofarads when the fluence exceeds $1 \times 10^{15} p/cm^2$, respectively as shown in figure 6 and figure 7(a) of [1]. GaAs photodiodes suffer from a relative responsivity loss over 40% at 850 nm wavelength when the fluence reaches $1 \times 10^{15} n/cm^2$, as shown in figure 4 of [1] and figure 4 of [2]. A higher reverse bias voltage for photodiodes may overcome these issues [3]. Based on these findings, we designed and prototyped a quad transimpedance and limiting amplifier ASIC named QTIA with fully differential architecture and an on-chip charge pump to research mitigation options.

## 2. Circuit Design

### 2.1 Overall structure

QTIA has four channels with a fully differential architecture. The block diagram of the QTIA is shown in figure 1. For each channel, a PIN Photodiode is AC coupled to the transimpedance amplifier (TIA) using on-chip capacitors. An on-chip biasing circuit provides proper biasing voltage for the photodiodes.

    A fully differential cascade TIA with programmable feedback resistance is designed to achieve low noise and adjustable bandwidth. The limiting amplifier (LA), comprising two stages with shared inductors and two third-order active feedback stages, has a total gain of over 40 dB. The active feedback cell induces two complex poles, which depend on the equivalent

feedback resistors, and achieves a flexible trade-off between the bandwidth and the gain peaking. AC coupling connection is applied between TIA and LA. The received signal strength indicator (RSSI), designed for optical alignment, is also implemented in this chip. The output driver drives an external 100 Ω differential load with adjustable amplitude. A DC offset-cancellation (DCOC) circuit feeds back the control voltage to the LA's second input stage. The power supplies of QTIA are 2.5 V and 1.2 V, separately, to save power. A generic I$^2$C module is used to configure the biasing currents and bandwidth.

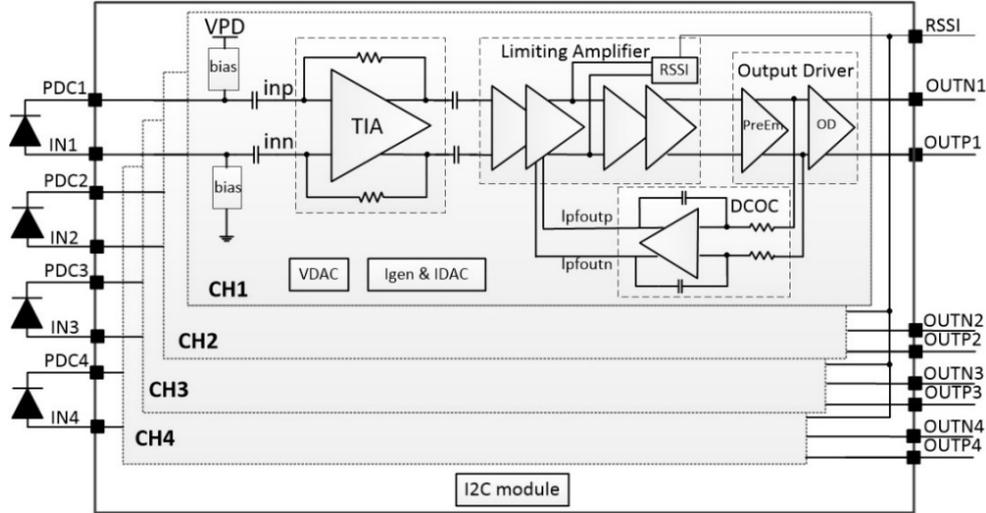

**Figure 1**: Overall structure of QTIA.

**2.2 Photodiode bias circuit**

The photodiode bias circuit is designed to maintain high AC impedance but relatively low DC impedance, as shown in figure 2(a). The Up-bias circuit is realized by a PMOS transistor with source degeneration and maintains a high AC impedance. The high AC impedance suppresses the photodiode bias voltage (VPD) noise and lowers the cut-off frequency with a reasonable size of the on-chip AC coupling capacitor. The Down-bias circuit comprises an NMOS transistor with source degeneration, and it has the equivalent AC impedance as the Up-bias circuit. The relatively low DC impedance helps with a proper voltage drop across the photodiodes.

In the prototype chip, the bias circuits in all channels are different for the convenience of studying the circuits' properties in the tests. Channel 2 only implements an Up-bias circuit, and the photodiode's anode (IN) is grounded. In contrast, Channel 3 implements a Down-bias circuit and directly connects the photodiode's cathode (PDC) to VPD. VPD is tied to a 2.5 V power supply in Channel 2 and Channel 3. Both Up-and Down-bias circuits are used in Channel 1 and Channel 4. A charge pump is implemented in Channel 1 to raise the photodiode bias voltage to be higher than 2.5 V, while the bias voltage is provided through an external power pad in Channel 4. Even with the low DC impedance, the radiation-induced increase in the dark current of InGaAs photodiodes results in a loss of the voltage drop across the photodiodes in such a way that the biasing voltage is insufficient. A higher bias voltage can mitigate the loss and allow for proper operation.

**2.3 Transimpedance Amplifier**

The differential TIA transfers an optical current to a differential voltage, as shown in figure 2(b). The feedback resistors are programmable to adjust the transimpedance gain ranging from 45 dB to 53 dB and the bandwidth between 3 GHz and 9 GHz, corresponding to 2.56 Gbps and 10

Gbps data rate options. In the case of a high loop-gain, the cascode structure is usually used to avoid the Miller effect [3]. Since the voltage amplifier has a relatively low gain (around 3) to favor the high bandwidth design for 10 Gbps data rate, it is not worth adding a cascode transistor to reduce the extra capacitance load due to the Miller effect. The power supply of TIA is chosen as 2.5 V to overcome the voltage drop of the resistors R1 and R2. Thus, the bias transistors (M3, M4, and M5) are added to ensure that the voltage across each transistor never exceeds 1.2 V for device reliability. The bias currents are also programmable for the different data rates.

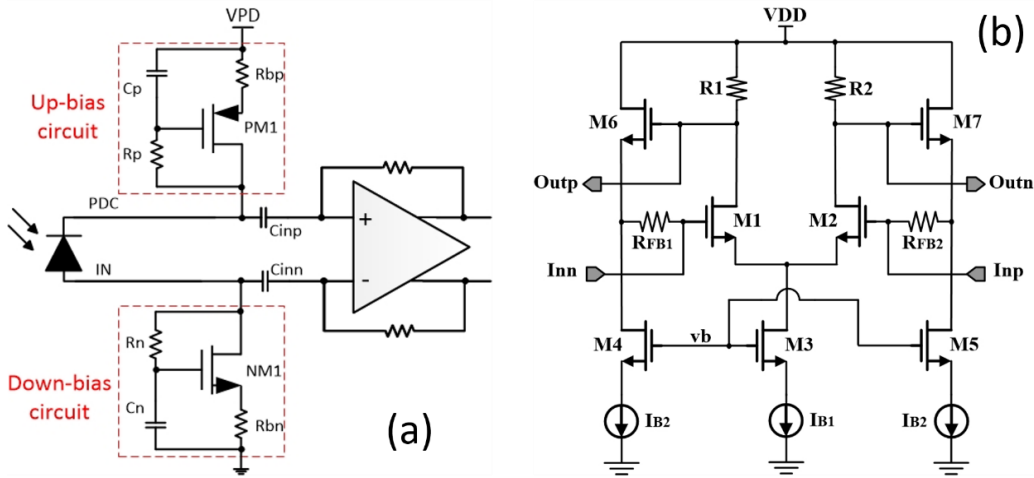

**Figure 2:** Schematic of the Photodiode bias circuit (a) and the transimpedance amplifier (b).

## 2.4 Charge pump

A cross-coupled charge pump is designed to raise the bias voltage to the dual bias circuit for the photodiode used in Channel 1. As shown in figure 3, the power supply of the charge pump is 2.5 V. To achieve the output voltage of over 5 V, three stages of voltage doublers with the dual-branch structure are employed [4]. A ring voltage-controlled oscillator (VCO), consisting of five inverters with a frequency of about 400 MHz, generates three pairs of clock signals (CLK and CLKB) to control the charging and discharging processes. The clock signals work in different phases ($\Delta T \approx 1$ ns) to decrease the transient current peak on the power supply and improve pumping efficiency. The reliability of all the transistors and capacitors is carefully verified to protect the circuits because the output voltage is much higher than 1.2 V, especially in the last two stages. In the simulation, the output voltage ranges from 5.4 V to 9.7 V, contributed by a load current from 1.25 mA (post-irradiation level) to 10 μA (pre-irradiation level) and the variations of process corner, power voltage, and temperature. The ripple voltage is below 10 mV.

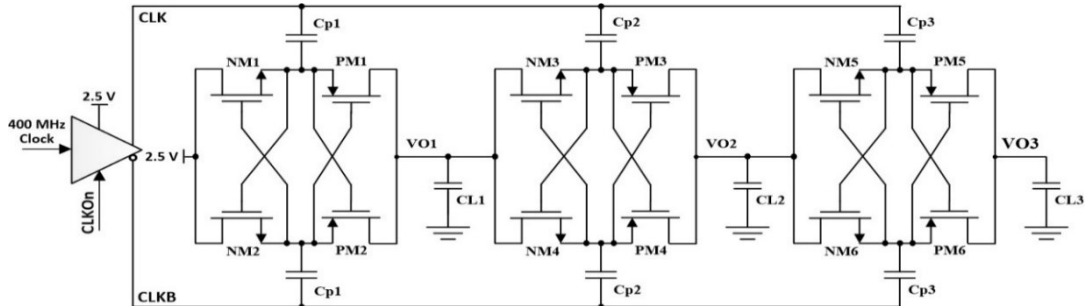

**Figure 3:** Schematic of the charge pump in Channel 1.

## 3. Test results

QTIA has been fabricated in a 65 nm CMOS process. The layout of QTIA is shown in figure 4 (a). The chip is 2 mm × 2 mm. To characterize the functions, QTIA is assembled in the ultra-small and lightweight optical module, QTRx, together with a 4-channel array laser driver named QLDD. QTIA was wire bonded to a high-speed array GaAs PIN Photodiode with a typical responsivity of 0.6 A/W and a parasitic capacitance of 90 fF.

Figure 4(b) presents the output eye diagram of Channel 4 measured at 2.56 Gbps for -6 dBm input power using a $2^7$-1 pseudo-random binary sequence (PRBS). The nominal differential output amplitudes remain constant at 400 mVpp, even for small input signals down to -18 dBm. The rise time is 40 ps, and the total jitter of 38.5 ps (around 0.1 Unit Interval) is combined by a random jitter of 1.8 ps and a deterministic jitter of 16.7 ps. The power consumption is below 120 mW and 72 mW, respectively, for Channel 1 with a charge pump and other channels without a charge pump. QTIA was also preliminary measured and performed at 10 Gbps. A well-open eye diagram of Channel 4 at 10 Gbps for -6 dBm input power was obtained in figure 4(c).

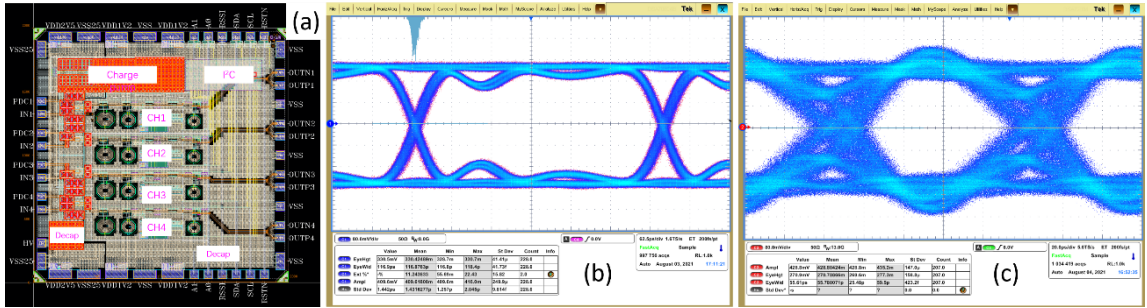

**Figure 4:** Layout of QTIA (a), output eye diagrams of Channel 4 for -6 dBm input power at 2.56 Gbps (b) and 10 Gbps (c).

QTIA performances at 2.56 Gbps versus input optical power are summarized in Figure 5. Channel 4 displays a sensitivity of -17 dBm for a bit error rate (BER) of 1E-12. As expected, the fully differential structure with photodiode bias circuits on both sides has an improved gain of about 3 dB. The on-chip charge pump in Channel 1 operates properly and provides comparable sensitivity and jitter performance with an external bias voltage of 2.5 V in Channel 4. The output swings and power consumptions are stable with different input power.

The output voltage of the charge pump in Channel 1 cannot be measured directly but estimated by the measured voltage on the PDC1 node as shown in figure 1. At room temperature, the PDC1 voltage ranges from 6 to 6.7 V corresponding to the input optical power between -1 dBm and -8 dBm. Accordingly, the output voltage of the charge pump without irradiation is estimated to be 6.5 V to 7 V, which is reasonable compared to the simulation.

At 2.56 Gbps and for -6 dBm input power, the effects of photodiode bias voltage were studied in Channel 4. As shown in Figure 6(a), the jitter performance can be optimized with a higher VPD (still below 2.5 V) due to the junction capacitance decrease and the bandwidth enhancement. Figure 6(b) illustrates the measured DC currents across the photodiodes, estimating half of input optical currents. The stable currents indicate that the pre-irradiated gain of responsivity is relatively small with a higher bias voltage for the tested GaAs Photodiode.

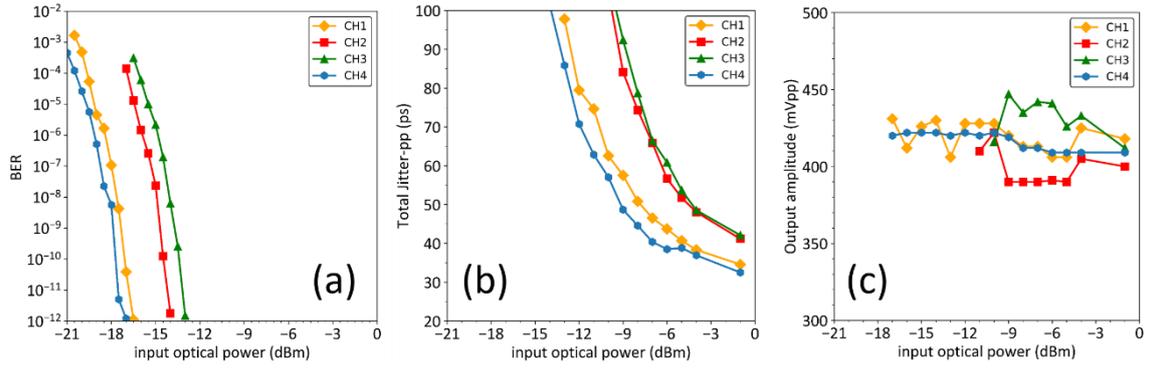

**Figure 5:** QTIA performance at 2.56 Gpbs: Sensitivity (a), total jitter (b), and output amplitude (c) versus input optical power.

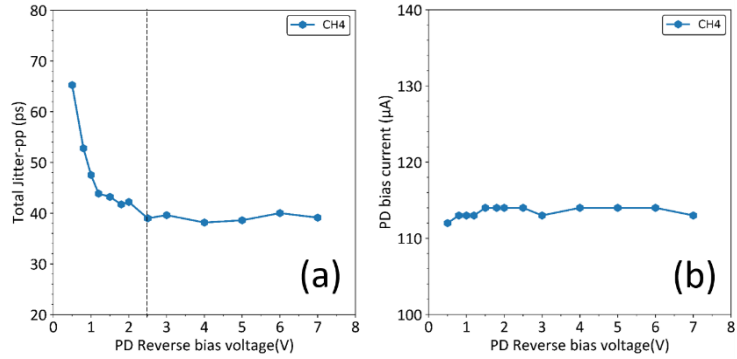

**Figure 6:** Total jitter (a) and DC currents across the photodiodes (b) versus PD bias voltage of Channel 4 at 2.56 Gbps.

## 4. Conclusion and Outlook

A 2.56 or 10 Gbps quad-channel transimpedance and limiting amplifier prototype chip has been developed to mitigate the irradiation-induced degradation of GaAs and InGaAs photodiodes. The circuit functions of this chip QTIA successfully passed the lab tests with GaAs photodiodes. As expected, the photodiode bias circuit on both sides (Up and Down) has better performance due to an extra gain of 3 dB. The on-chip charge pump used in Channel 1 meets expectations. This channel provides comparable sensitivity and jitter performance with the reference channel with an external bias voltage. A higher photodiode bias voltage leads to better jitter performances, possibly due to a reduced junction capacitance.

The benefits of an on-chip charge pump in radiations can be foreseen with both GaAs and InGaAs photodiodes. The on-chip charge pump ensures that the optimal voltage across photodiodes can be guaranteed, and the dual bias circuit for photodiodes can be adopted, even with a large dark current of InGaAs photodiodes after irradiation. The effects of a higher bias voltage on the responsivity during irradiation are to be studied. A full set of irradiation tests with different photodiodes will be arranged to further verify the benefits of the on-chip charge pump in QTIA.

## Acknowledgments

This work is supported by SMU's Dedman Dean's Research Council Grant and the National Science Council in Taiwan.